%% file: main.tex
\documentclass[runningheads]{llncs}
\usepackage{graphicx}
\usepackage{titling}
\usepackage{amsmath}
\usepackage{amsfonts}
\usepackage{enumerate}
\usepackage{xcolor}

\usepackage{hyperref}

\hypersetup{
  hidelinks,
  colorlinks=true,
  allcolors=black,
  pdfstartview=Fit,
  breaklinks=true
}

\usepackage[all]{hypcap}

\usepackage[caption=false,font=footnotesize]{subfig}

\usepackage{mindflow}

\usepackage[capitalise,nameinlink]{cleveref}

\crefname{section}{Sect.}{Sect.}
\Crefname{section}{Section}{Sections}
\crefname{listing}{List.}{List.}
\crefname{listing}{Listing}{Listings}
\Crefname{listing}{Listing}{Listings}
\crefname{lstlisting}{Listing}{Listings}
\Crefname{lstlisting}{Listing}{Listings}

\usepackage{algorithm2e}
\RestyleAlgo{ruled}
\SetKwComment{Comment}{/* }{ */}

\newcommand{\bpf}{\textit{Proof.} }
\newcommand{\epf}{\hfill$\square$}
\DeclareMathOperator{\vcdim}{VC-dim}
\DeclareMathOperator{\cardinality}{card}
\newcommand{\card}[1]
{\cardinality\hspace{-.2em}\left({#1}\right)}
\DeclareMathOperator{\length}{len}
\newcommand{\len}[1]
{\length\hspace{-.2em}\left({#1}\right)}
\newcommand{\sets}{\mathcal{R}}
\newcommand{\cells}[1]{\varphi\left(#1\right)}
\newcommand{\reduce}[2]{{#1\rvert}_{#2}}
\newcommand{\sdiff}[2]{\Delta(#1, #2)}
\newcommand{\bigO}[1]{\mathcal{O}\left(#1\right)}
\newcommand{\Ex}{\mathbb{E}}
\renewcommand{\Pr}{\mathbb{P}}
\newcommand{\e}{\epsilon}
\newcommand{\pack}{\mathcal{P}}
\newcommand{\measure}[1]{\mu\left(#1\right)}

\newcommand{\mcS}{\mathcal{S}}
\newcommand{\mcT}{\mathcal{T}}
\newcommand{\nsets}{n}
\newcommand{\neles}{m}
\newcommand{\npack}{p}

\newcommand{\newreptheorem}[2]{%
\newenvironment{rep#1}[1]{%
 \def\rep@title{#2 \ref{##1}}%
 \begin{rep@theorem}}%
 {\end{rep@theorem}}}

\begin{document}
\title{Hitting Sets when the Shallow Cell Complexity is Small
\thanks{This material is based on work supported by the NSF under Grant CNS-1952063.}}
%
%
\author{Sander  Aarts\inst{1}\orcidID{0000-0003-1852-9116} \and
David B. Shmoys\inst{1}\orcidID{0000-0003-3882-901X}
}
\authorrunning{S. Aarts,  D.B. Shmoys}
%

\institute{Cornell University, Ithaca NY 14580, USA\\
\email{sea78@cornell.edu, david.shmoys@cornell.edu}}
\maketitle              
\begin{abstract}
The hitting set problem is a well-known NP-hard optimization problem in which, given a set of elements and a collection of subsets, the goal is to find the smallest selection of elements, such that each subset contains at least one element in the selection. Many geometric set systems enjoy improved approximation ratios, which have recently been shown to be tight with respect to the shallow cell complexity of the set system. The algorithms that exploit the cell complexity, however, tend to be involved and computationally intensive. This paper shows that a slightly improved asymptotic approximation ratio for the hitting set problem can be attained using a much simpler algorithm: solve the linear programming relaxation, take one initial random sample from the set of elements with probabilities proportional to the LP-solution, and, while there is an unhit set, take an additional sample from it proportional to the LP-solution. Our algorithm is a simple generalization of the elegant net-finder algorithm by Nabil Mustafa. To analyze this algorithm for the hitting set problem, we generalize the classic Packing Lemma, and the more recent Shallow Packing Lemma, to the setting of weighted epsilon-nets.

\end{abstract}
%
%
%

\section{Introduction}
\label{sec:intro}
The input to the hitting set problem is a finite \emph{set system} -- a ground set $X$ of $\neles$ elements, or \textit{points}, and a collection $\sets$ of $\nsets$ subsets, or \textit{ranges}, of $X$. This can also be understood as a hypergraph, with vertices $X$ and hyper-edges $\sets$. A \emph{hitting set} is a subset of elements $H \subseteq X$ such that every set $R \in \sets$ is hit by $H$, i.e. $R \cap H \neq \emptyset$, for all $R \in \sets$. This is a vertex cover under the hypergraph view.
The set system can be encoded as a set-element incidence matrix $A \in \{0, 1\}^{\nsets \times \neles}$, in which the $(i,j)$th entry $a_{ij}$ is $1$ if range $R_i$ contains point $x_j$, and $0$ otherwise. The IP of the minimum hitting set problem is
\begin{align}
    \label{eq:ip_formulation}
    \min_y \sum_{j: x_j \in X}&y_j \nonumber\\ 
    \textrm{s.t. } \sum_{j: x_j \in X} &a_{ij}y_j \geq 1, && \forall i : R_i \in \sets;\\
    & y_j \in \{0, 1\}, && \forall j: x_j \in X,\nonumber
\end{align}
where variable $y_j \in \{0, 1\}$ indicates whether element $x_j$ is in the solution $H$. 

Hitting sets and set covers are intimately connected; a hitting set for $A$ is a set cover of $A^T$. Both problems' decision versions are NP-complete \cite{garey1979computers}. There exists an $\bigO{\log \neles}$-approximation algorithm, and this bound is tight unless P = NP \cite{feige1998threshold,johnson1973approximation}. However, there are algorithms that exploit additional structure in $A$ to attain improved approximation ratios\footnote{For example when $A$ has bounded row or column sums \cite{bar1981linear,chvatal1979greedy}.}. Indeed, our work is motivated by the problem of exploiting structure when covering large numbers of wireless LoRaWAN transmitters with wireless receivers. Transmitters can be viewed as points, which are considered to be covered if they are in the line of sight of a wireless receiver, which in turn drives transmission quality in LoRaWAN \cite{YousufLoRa}. The area in the line of sight of a receiver roughly resembles a simple shape.

Many geometric set systems enjoy better approximation ratios via \emph{epsilon-nets}, or $\e$-nets. A set system is said to be \emph{geometric} whenever its elements can be encoded as points in Euclidean space, and sets are derived from containment of the points in geometric shapes, such as half-spaces, balls or rectangles\footnote{Some definition allow for uncountably many geometric shapes in $\sets$, e.g. all squares. However, because the number of points $X$ is finite, there are nevertheless a finite number of unique sets induced by these shapes.}. The seminal work of Brönnimann and Goodrich \cite{bronnimann1995almost}, and Even \textit{et al.} \cite{even2005hitting}, connects the approximability of a hitting set instance to the size of weighted $\e$-nets.  Given non-negative weights on the points, $\mu: X \rightarrow \mathbb{R}_{\geq 0}$, a \emph{weighted} $\e$-net with respect to weights $\mu$ is a subset $H \subseteq X$ that hits all $\e$-heavy sets:
\begin{equation}
    \label{eq:e-net}
    \forall R \in \sets \textrm{ with } \mu(R) \geq \e \cdot \mu(X): \quad R \cap H \neq \emptyset,
\end{equation}
where the weight of any subset $S \subseteq X$ is defined as $\mu(S) = \sum_{x \in S}\mu(x)$. Even \textit{et al.} \cite{even2005hitting} reduce the problem of finding a small hitting set to finding a small $\e$-net via a reformulation of the linear programming relaxation of the hitting set problem (\ref{eq:ip_formulation}). The reformulated LP (\ref{eq:even_lp_formulation}) is a program for finding the largest $\epsilon$, and corresponding weights $\mu$, subject to the constraint that an $\e$-net with respect to weights $\mu$ is a hitting set.
\begin{align}
\label{eq:even_lp_formulation}
    \max_{\e, \mu} \ &\epsilon \nonumber\\ 
    \textrm{s.t. } \sum_{j: x_j \in X} a_{ij}&\mu_j \geq \e, && \forall i: R_i \in \sets;\nonumber\\
     \sum_{j: x_j \in X} &\mu_j = 1;\\
    &\mu_j \geq 0,  && \forall j :  x_j \in X.\nonumber
\end{align}
The first constraint requires that each set $R$ is $\e$-heavy; the second constraint normalizes the weights.   Let $(\epsilon^*, \mu^*)$ denote an optimal solution to LP (\ref{eq:even_lp_formulation}), with $\mu^* = (\mu^*_1, \dots, \mu^*_n)$. Let $z^*$ be the optimal value to the LP relaxation of the original program (\ref{eq:ip_formulation}). The first constraint ensures that an $\e^*$-net with respect to weights $\mu^*$ is a hitting set. Moreover, the reciprocal optimal value $1/\e^*$ is equal to the optimal LP value $z^*$ \cite{even2005hitting}. In particular, an $\e^*$-net of size $g(1/\e^*)$ for some function $g(\cdot)$ is a hitting set of size of $g(z^*)$. Hence, to find a small hitting set it suffices to solve LP (\ref{eq:even_lp_formulation}) and find a small $\e^*$-net with respect to weights $\mu^*$. 


Haussler and Welzl \cite{haussler1986epsilon} show that set systems with bounded VC-dimension admit small $\e$-nets, and develop a simple algorithm to find them. The VC-dimension is a measure of the set system's complexity. Given a subset $S \subseteq X$, the \emph{projection} of $\sets$ to $S$ is the set system formed by elements $S$ and sets $\reduce{\sets}{S} = \{R \cap S: R \in \sets\}$. The VC-dimension of $\sets$ is the size of the largest subset $S \subseteq X$ such that $\reduce{\sets}{S}$ \emph{shatters} $S$, i.e. the largest set $S$ such that $\reduce{\sets}{S}$ contains all subsets of $S$. In particular, Clarkson \cite{clarkson1988randomized}, and Haussler and Welzl \cite{haussler1986epsilon}, show that any set system with VC-dimension $d$ has a weighted $\e$-net of size  $\bigO{\tfrac{d}{\e}\log\tfrac{1}{\e}}$. This is remarkable, as the size is independent of both the size of $X$ and $\sets$. 
Moreover, the algorithm for finding such  an $\e$-net is simple: Select a subset $H \subseteq X$ by sampling each element $x$ in $X$ independently. 
\begin{theorem}[$\e$-net Theorem \cite{haussler1986epsilon,KomlosPW92}]
\label{thm:e-net}
    Let $(X, \sets)$ be a set system with VC-dimension $d$, and let $\mu: X \rightarrow \mathbb{R}_{\geq 0}$ be element weights with $\mu(X) = 1$. Then for any $\e, \gamma \in (0, 1)$:
    \begin{equation*}
        H \gets \textrm{ pick each } x \in X \text{ with probability  }\min\left\{1, \frac{2\mu(x)}{\e}\cdot\max\left\{\log\tfrac{1}{\gamma}, d\log\tfrac{1}{\e}\right\}\right\}
    \end{equation*}
is a weighted $\e$-net with respect to weights $\mu$ with probability at least $1-\gamma$.
\end{theorem}
Throughout, we define $\mu(S) = \sum_{x \in S}\mu(x)$ for all subsets $S \subseteq X$. For general set systems of VC-dimension $d$, this bound is tight in expectation \cite{KomlosPW92}. However, there are alternative ways to parameterize the complexity of set systems.

\subsection{Shallow Cell Complexity}
The \textit{shallow cell complexity} (SCC) is a finer parameterization of the  complexity of set systems.
\cite{aronov2009small,ChanQuS,varadarajan2009epsilon}.
Readers are referred to Mustafa and Varadarajan \cite{mustafa2017epsilon} for more background.  A \emph{cell} in a binary matrix $A$ is a collection of identical rows. A cell has \emph{depth} $k$ if the number of $1$'s in any of its rows is exactly $k$, i.e., if each set in the cell contains $k$ elements. For a non-decreasing function $\cells{\cdot, \cdot}$ we say binary matrix $A$ has \emph{shallow cell complexity} (SCC) $\cells{\cdot, \cdot}$ if, for all $1 \leq k \leq l \leq \neles$, the number of cells of depth at most $k$ in any submatrix $A^*$ of $A$ of at most $l$ columns, is at most $\cells{l, k}$. A set system $(X, \sets)$ is said to have SCC $\cells{l, k}$ if its set-element incidence matrix $A$ does. Often $\cells{l, k} = \bigO{\cells{l}k^c}$ for some constant $c > 0$ and single-variable function $\cells{\cdot}$, in which case the dependence on $k$ is can be dropped and the SCC denoted by $\cells{l}$. Examples of geometric set systems with small shallow cell complexity are discs in the plane with $\cells{l, k} = \bigO{k}$, and axis-parallel rectangles with $\cells{l, k} = \bigO{lk^2}$ \cite{mustafa2022sampling}. 

As is true for VC-dimension, there are algorithms that find hitting sets or $\e$-nets with sizes bounded in terms of the shallow cell complexity. A prominent example is the quasi-uniform sampling algorithm of Chan \textit{et al.} \cite{ChanQuS}. Given non-negative weights $\mu: X \rightarrow \mathbb{R}_{\geq 0}$, and a value $\epsilon > 0$, the algorithm finds a hitting set while maintaining an upper bound on the probability of selecting any given element.
\begin{theorem}[Quasi-uniform sampling \cite{ChanQuS}]
    Suppose a set system defined by $A$ has SCC $\cells{l, k} = \cells{l}k^c$ for some $c >0$. Then there is a randomized poly-time algorithm that returns a hitting set of expected size $\bigO{\max\{1, \log(\cells{\neles})\}}$ times the LP optimum.
\end{theorem}
The algorithm attains the optimal approximation ratio with respect to the SCC\footnote{In addition, it is worth noting that this algorithm can solve the more general \textit{weighted} hitting set problem, in which each element has a given weight, and the goal is to find the minimum weight hitting set.}. However, the sampling procedure is involved, and may require enumeration over all sets $\sets$, of which there can be $\nsets = \Omega(\neles^c)$ for some constant $c > 0$ \cite{mustafa19}. 

Taking a different approach, 
Mustafa and colleagues \cite{Mustafa16,mustafa19,mustafa2018simple} develop a net-finder for asymptotically optimal-sized \emph{un}weighted $\e$-nets with respect to the SCC. The algorithm is remarkably simple: Take an initial sample from $X$, and while there are unhit sets, choose an unhit set arbitrarily, and add $\bigO{1}$ randomly chosen elements from this set to the original sample. The algorithm assumes access to an oracle that returns an unhit set. This oracle is called at most $\bigO{1/\e}$ times in expectation. While the size of the returned $\e$-net is asymptotically on par with the quasi-uniform sampling algorithm, there are large constants in the upper bound \cite{mustafa19}.

This algorithm is not directly applicable to the hitting set problem via the LP-reduction above, although it can be used via a standard reduction. The analysis of the algorithm applies to only uniform weights, and the optimal weights $\mu^*$ of the LP-formulation (\ref{eq:even_lp_formulation}) are not generally uniform. Nevertheless, it is possible to reduce the problem of finding a weighted $\e$-net to that of finding a uniform $\e'$-net following a standard reduction, in which an expanded instance is generated by copying each element $x_j \in X$ a number of times roughly proportional to its weight $\mu^*(x_j)$ \cite{bronnimann1995almost,ChanQuS}. This can generate $\Omega(\neles)$ copies of each element, which can have notable consequences. First, to achieve a weighted $\e^*$-net in the original instance, one must use a smaller value $\e'$ for the expanded instance, on the order of $\bigO{\e^* / \neles}$. This results in an approximation ratio  of $\bigO{\log\cells{\bigO{\neles}}}$. Secondly, generating copies can increase the number of elements from $\neles$ to $\Omega(\neles^2)$. This can increase the runtime considerably. In particular, repeatedly sampling from sets of size $\Theta(\neles^2)$ can become prohibitive on large instances such as the wireless coverage problem motivating our work.

\subsection{Our Contributions}
This paper generalizes the elegant net-finder algorithm of Mustafa \cite{mustafa19} to the setting of weighted $\e$-nets, in order to produce a fast and simple algorithm for the hitting set problem, which attains asymptotically optimal approximation ratios with respect to the shallow cell complexity. The algorithm enjoys a faster runtime that makes solving larger instances, such as LoRaWAN receiver placement at scale, feasible. This is achieved by combining the weighted $\e$-net finder with the reduction of Even \textit{et al.} \cite{even2005hitting}. In doing so, we also improve on the asymptotic approximation ratio from $\max\{1, \log\cells{\neles}\}$ to $\max\{1, \bigO{\log\cells{\bigO{z^*}}}\}$ where $z*$ is the optimal value to the linear relaxation of the hitting set program (\ref{eq:ip_formulation}). While in the worst case $z^* = \neles$, it is often the case that $z^* \ll \neles$. However, the multiplicative constants in our analysis are relatively large, matching those of Mustafa \cite{mustafa19}. In addition to the algorithm, our analysis generalizes the classic Packing Lemma of Haussler \cite{Haussler1995ShperePacking}, as well as the Shallow Packing Lemma of Mustafa \textit{et al.}. \cite{mustafa2018simple}, to the weighted setting, which may be of independent interest.

Key to our approach are adaptations of Mustafa's \cite{mustafa2018simple} Shallow Packing Lemma and Haussler's \cite{Haussler1995ShperePacking} classic Packing Lemma that accommodate non-uniform weights. Our main technical contribution is to allow a notion of \emph{weighted packings}. Consider any non-negative weights $\mu: X \rightarrow \mathbb{R}_{\geq 0}$ with $\sum_{x \in X}\mu(X) = 1$, and extend it to element subsets via $\mu(S) = \sum_{x \in S}\mu(S)$.\footnote{Any non-negative weights $w:X \rightarrow \mathbb{R}_{\geq 0}$ with $w(X) > 0$ can be normalized as $\mu(x) = w(x)/w(X)$.} A $(k, \delta)$-\emph{packing with respect to weights $\mu$} is a collection of sets $\pack \subseteq \sets$ in which (i) all sets $R$ in $\pack$ are at most $k$-\textit{heavy}, i.e., have bounded weight $\mu(R) \leq k$; and (ii) all pairs of sets have symmetric differences of weight at least $\delta$. (See Definition \ref{def:k-pack}). Our weighted shallow packing lemma upper bounds the number of sets in $\pack$ as a function of the SCC. Our approach accommodates weights $\mu$ by sampling elements from a distribution with probability mass proportional to the weights, rather than from a uniform distribution as in the original proofs. Moreover, our proof uses sampling \textit{with replacement} rather than \textit{without replacement} to simplify the analysis. While more generally applicable, our result yields the same bound on the size of $\pack$ as in the unweighted setting. An analogous sampling approach is used in proving \cref{thm:e-net} \cite{KomlosPW92}.  Equipped with our generalized lemma, it is straightforward to adapt Mustafa's \cite{mustafa19} analysis to a weighted net-finder. A proof of our Weighted Packing Lemma is included in the extended online version.


\section{Algorithm and Main Result}
\label{sec:algo_and_results}

Our algorithm combines the LP-relaxation of Even \textit{et al.} \cite{even2005hitting} with the generalized sampling approach of Mustafa \cite{mustafa19}. Our procedure is summarized in Algorithm \ref{alg:detailed}.
The algorithm makes use of two global constants, $\beta$ and $\gamma$. These are assumed to be positive, and to satisfy $\gamma \leq 1/4$ and $\beta + \gamma \leq 1$.
\begin{algorithm}[ht]
\caption{A simple hitting set algorithm with details}\label{alg:detailed}
\KwData{A matrix $A$ with $\vcdim(A) \leq d$ and SCC $\cells{\cdot, \cdot}$, constants $\gamma, \beta > 0$}
$\e^*, (\mu^*_1, \dots, \mu^*_n) \gets$ solve LP  $\{\max \e: A\mu \geq \e, \mu^T \mathbf{1} = 1, \mu \geq \mathbf{0}\}$\;
$H \gets \emptyset$\;
\For{$x_j \in X$}{
    $H \gets H \cup \{x_j\}$ with probability $\min\left\{1, \dfrac{2\mu^*_j}{\left(\tfrac{3}{4} - \tfrac{\beta}{2}\right)\e^*}\cdot \max\left\{\log\left(d^2\cells{\tfrac{8d}{\beta\e^*}, \tfrac{48d}{\beta}}^2\right), d\log\left(\dfrac{1}{\left(\tfrac{3}{4}- \tfrac{\beta}{2}\right)\e^*}\right)\right\} \right\}$
  }
\While{there is a set $R \in \sets$ not hit by $H$}{
Independently add each $x_j \in R$ to $H$ with probability $\min\left\{ 1, \frac{2\mu^*_j}{\gamma \mu^*(R)}\max\{\log 2, d\log\tfrac{1}{\gamma}\}\right\}$
}
\Return{$H$}
\end{algorithm}%

In the while loop, the weights $\mu^*(R) = \sum_{j: x_j \in R}\mu^*_j$ denote the weight of set $R$ under the LP optimal weights $\mu^* = (\mu^*_1, \dots, \mu^*_n)$.

Conceptually, the algorithm is simple; it randomly selects an initial set of elements $H$ from $X$, and proceeds to add additional random subsets of elements to $H$ until this is a hitting set. The algorithm relies on an oracle that returns an arbitrary unhit set. This oracle is treated as a black box. Our main result is twofold: we bound the expected size of the solution hitting set $H$ as a function of the cell complexity, and bound the expected number of oracle calls.
\begin{theorem}
\label{thm:algo_with_consts}
Let $A$ be a binary matrix encoding a hitting set instance with shallow cell complexity $\cells{\cdot, \cdot}$ and $\vcdim(\sets) \leq d$. Let $z^*$ be LP optimal value. Then the algorithm returns a hitting set of expected size
\begin{equation*}
    \bigO{z^* \cdot \max\left\{{1, \log\cells{\bigO{z^*}, \bigO{d}}}\right\}}.
\end{equation*}
Furthermore it makes $\bigO{z^*}$ oracle calls in expectation.
\end{theorem}
Note that the algorithm always returns a hitting set; the randomness is in the size of the solution and the runtime. This is in contrast with the net-finder in \cref{thm:e-net}. Both algorithms require knowing the VC dimension $d$; ours must additionally know the shallow cell complexity $\cells{\cdot, \cdot}$. If unknown, these can be searched for using a standard doubling trick \cite{mustafa19}.



\section{The Weighted Shallow Packing Lemma}
\label{sec:packing}

The Weighted Shallow Packing Lemma is key to proving \cref{thm:algo_with_consts}. This section formally defines weighted shallow packings, states the lemma, and proves it. To this end, fix non-negative weights $\mu$ over $X$, and define the weight of a subset of elements $S \subseteq X$ as $\mu(S) = \sum_{j \in S}\mu_j$. Assume that $\mu(X) = 1$. To contrast, let $\card{S}$ denote the cardinality of any set $S$. Note that the weights $\mu$ induce a probability distribution over the elements $X$. Throughout, whenever an element $u$ of $X$ is randomly sampled, it is assumed to follow a distribution proportional to $\mu(\cdot)$, in which case we say $u$ is sampled from $\mu(\cdot)$, and denote this by $u \sim \mu(\cdot)$. Note that an element $u \sim \mu(\cdot)$ sampled this way lies in subset $S \subseteq X$ with probability $\mu(S)$.


The main purpose of the weighted shallow packing lemma is to bound the number of sets in a set system in terms of its shallow cell complexity. Clearly, an arbitrary set systems can contain large numbers of sets. Instead, we focus on a particular kind of set system called a \emph{weighted packing}. A set system is a packing if all its sets are ``light'', and each pair of sets are sufficiently different from each other. Critically, we define ``light'' and ``different'' in reference to the weights.
\begin{definition}
    \label{def:k-pack}
    Let $(X, \pack)$ be a set system with weights $\mu$, and let $k, \delta \in (0, 1)$ be constants. If all sets $S$ in $\pack$ satisfy $\mu(S) \leq k$, and all pairs of distinct sets $S, R$ in $\pack$ have symmetric difference of weight at least $\delta$, i.e.
\begin{equation}
    \label{eq:symmetric_difference}
    \measure{\sdiff{S}{R}} = \measure{(S \backslash R) \cup (R\backslash S)} \geq \delta,
\end{equation}
then we say $(X, \mathcal{P})$ is a weighted $(k, \delta)$-packing with respect to $\mu$.
\end{definition}
We omit the ``with respect to $\mu$''-statement whenever this is clear from context.

The shallow packing lemma bounds the number of sets in a packing as a function of the constants $(k, \delta)$, the VC-dimension, and the shallow cell complexity.
\begin{lemma}[Weighted shallow packing lemma]
\label{lem:shallow_packing}
Let $(X, \pack)$ be a set system on $\neles$ elements, equipped with weights $\mu$, and let $(X, \pack)$ be a $(k, \delta)$-packing with respect to $\mu$ for constants $k, \delta > 0$. Assume the set system has $\vcdim(\pack) \leq d$, and shallow cell complexity $\cells{\cdot, \cdot}$. Then
\begin{equation*}
    \card{\pack} \leq \frac{24d}{\delta} \cdot  \cells{\frac{8d}{\delta}, \frac{48dk}{\delta}}.
\end{equation*}
\end{lemma}
The proof to this lemma makes use of our weighted Packing Lemma. The unweighted Packing Lemma is a classic result by Haussler \cite{Haussler1995ShperePacking} that bounds the number of sets in a packing. We generalize this to nonuniform weights.
\begin{lemma}[Weighted packing lemma]
\label{lem:packing}
Let $(X, \pack)$ be a set system with $\nsets$ sets and $\neles$ elements, equipped with weights $\mu$. Let $\vcdim(\pack) \leq d$ for some integer $d \geq 1$ and assume there is a constant $\delta \in (0, 1)$ such that $\mu(\Delta(S_i, S_k)) \geq \delta$ for all $1 \leq i < k \leq \nsets$. Then
\begin{equation*}
    \card{\pack}  \leq 2\Ex \left[\card{\reduce{\pack}{Y}}\right],
\end{equation*}
where $Y$ is the set of unique elements in a random sample $U = (u_1, u_2, \dots, u_s)$ of size $s = \lceil \frac{8d}{\delta}\rceil -1$, in which each element $u_k$  is sampled iid $u_k \sim \mu(\cdot)$ with replacement. 
\end{lemma}
The proof of the latter lemma is in the appendix to the 
extended online version of the paper; \cref{lem:shallow_packing} is proved next.

\subsection{Proof of the Weighted Shallow Packing Lemma}
\begin{proof}
 Fix a $(k, \delta)$-packing $\pack$ and let $U = (u_1, u_2, \dots, u_s)$ be a random sample of length $s$, in which each element is sampled $u_k \sim \mu(\cdot)$, $k=1, \dots, s$, independently and with replacement. The number of elements sampled is $s = \lceil \frac{8d}{\delta}\rceil - 1$. Let $Y \subseteq X$ be the set of unique elements in $U$. For every set $R \in \pack$, let $M(R, U) := \sum^s_{i=k}\mathbf{1}[u_k \in R]$ denote number of (copies of) elements in $U$ that are in $R$.
 Define $\pack_L \subseteq \pack$ as the sub-collection of ``large'' sets in packing $\pack$ that contain at least $6\left(\frac{8dk}{\delta}\right)$ (copies of) elements in the random sample $U$:
\begin{equation*}
    \mathcal{P}_L = \left\{R \in \pack: M(R, U) \geq 6 \cdot \frac{8dk}{\delta}\right\}.
\end{equation*}
It follows that the probability of a given range $R$ in $\pack$ being a member of $\pack_L$ is
\begin{equation*}
    \Pr[R \in \pack_L] = \Pr\left[M(R, U) \geq 6 \cdot\frac{8dk}{\delta}\right].
\end{equation*}
Our goal is to show that the collection of large sets $\pack_L$ has few members in expectation. To do so, it suffices to bound the probability that a fixed set $R$ is a member of $\pack_L$. This is achieved using Markov's inequality. Recalling that all sets $R \in \pack$ have bounded weight $\mu(R) \leq k$ gives
\begin{equation*}
    \Ex[M(R, U)] =  \sum^{s}_{k=1}\Pr[u_k \in R] = \sum^s_{k=1}\mu(R) \leq s \cdot k \leq  \frac{8dk}{\delta},
\end{equation*}
where we used the fact that we sample from $\mu(\cdot)$, which implies that $\Pr[u_k \in R] = \mu(R)$.
Now, Markov's inequality bounds the probability of $R$ being in $\pack_L$:
\begin{align*}
    \Pr[R \in \pack_L] &= \Pr\left[M(R, U) \geq 6 \cdot \frac{8dk}{\delta}\right] \\
    &\leq\Pr\Big[M(R, U) \geq 6\cdot \Ex[M(R, U)]\Big] \leq 1/6.
\end{align*}
Finally, because $\pack_L \subseteq \pack$, we conclude that
\begin{align*}
    \Ex[\card{\reduce{\pack}{Y}}] 
    &\leq \Ex[\card{\pack_L}] + \Ex[\card{\reduce{(\pack \backslash \pack_L)}{Y}}] \\
    &\leq \sum_{R \in \pack}\Pr[R \in \pack_L] + \card{Y}\cdot \cells{\card{Y}, 6 \cdot \frac{8dk}{\delta}}\\
    &\leq \tfrac{1}{6}\card{\pack} + \tfrac{8d}{\delta}\cdot \cells{\tfrac{8d}{\delta}, \tfrac{48dk}{\delta}},
\end{align*}
where the second-to-last inequality uses the shallow cell complexity of $\pack$; the system $(Y, \reduce{(\pack \backslash \pack_L)}{Y})$ has at most $\card{Y} \leq s$ elements, and sets have depth at most $\left(6\cdot \tfrac{8dk}{\delta}\right)$, as the system consists only of cells that are not ``large''. The final inequality holds because $\Pr[R \in \pack_L] \leq 1/6$. Finally, applying \cref{lem:packing} completes the proof. \epf 
\end{proof}

\section{Proof of the Main Theorem}
\label{sec:main_proof}
Equipped with the Weighted Shallow Packing Lemma, we follow a similar strategy as Mustafa \cite{mustafa19}. We state and prove three key lemmas, and finally prove \cref{thm:algo_with_consts}. 

\subsection{Key lemmas}
The proof of our main theorem relies on all sets having similar weight. Let $\e$ and $\mu = (\mu_1, \dots, \mu_n)$ be a feasible solution to the LP relaxation (\ref{eq:even_lp_formulation}). By the constraints of the LP, each set $R \in \sets$ has weight $\mu(R) = \sum_{j:x_j \in R}\mu_j \geq \e$. Partition the collection of sets $\sets$ into groups $\ell = 0, 1, \dots, \lceil \log \e \rceil$ of sets of similar weight; set $R$ belongs to group $\ell$ if and only if $2^{-\ell - 1}\e \leq \mu(R) < 2^{-\ell}\e$. Because the algorithm exclusively takes independent samples, we can view one run of the algorithm as multiple parallel, independent runs on each group of sets. All our bounds scale on the order $\bigO{1/ (2^{-\ell}\e)}$, so summing over the groups gives a final bound on the order of $\bigO{1/ \e}$. Hence, we assume henceforth that all sets $R \in \sets$ have weight $\e \leq \measure{R} \leq 2\e$.

The key idea of the proof is to amortize the elements added from each processed unhit set throughout the run of the algorithm. We say a set is \emph{processed} each time it is flagged as unhit by the oracle, and a sample is taken from it. We bound the total number of elements sampled using weighted $(k, \delta)$-packings on two levels. The first-level packing is an arbitrary \emph{maximal} packing $\pack$ of sets in $\sets$. There are a bounded number of sets in $\pack$. Next, each processed set  $R_i$ is assigned to a set in the first-level packing $\pack$. For a fixed set $P^j$ in the first-level packing, \emph{given that it has been assigned processed sets}, we show that the collection of sets $R_i$ assigned to $P^j$ forms a second-level packing. Each second-level packing also has a bounded number of sets. Finally, by bounding the probability that a set in the first-level packing has any sets assigned to it, the total expected number of times the algorithm processes a set is bounded. Note that the assignments are only a tool for analysis; they need not be computed by the algorithm. 

We begin by defining the first-level packing. Fix a \emph{maximal} $(2\e, \beta \e)$-packing $\pack = \{P^1, \dots, P^{\npack}\}$, where $p$ denotes the number of sets in the packing. The Shallow Packing \cref{lem:shallow_packing} upper bounds the number of sets in the packing by
\begin{equation}
\label{eq:size_pack}
    \npack \leq \frac{24d}{\beta \e} \cdot \cells{\frac{8d}{\beta\e}, \frac{96d}{\beta}}.
\end{equation}
Now, suppose the algorithm runs for $T$ steps, processing sets $(R_1, \dots, R_T)$ in sequence. One given set may be processed multiple times. Denote the sets of sampled elements $H_{R_1}, \dots, H_{R_T}$. The processed sets $R_i$ are assigned to sets $P^j$ in the first-level packing $\pack$ as follows. Arbitrarily assign each set $R_i$ to any index $j \in \{1, 2, \dots, \npack\}$ satisfying $\mu\left(\sdiff{R_i}{P^j}\right) < \beta \e$. Such an index $j$ exists because $\pack$ is a maximal $(2\e, \beta\e)$-packing. It may be the case that $R_i = P^j$. The next task is to bound the number of sets $R_i$ assigned to any set $P^j$ in the first-level packing.

Let $n_j$ denote the number of processed sets in $(R_1, \dots, R_T)$ assigned to $P^j \in \pack$. For now, condition on first-level packing set $P^j$ having at least one set assigned to it, i.e. $n_j \geq 1$. We study the probability of this event later. Relabel the sets and consider them in the order in which they were processed by the algorithm,
\begin{equation*}
    \mcS^j = (R^j_1, \dots, R^j_{n_j}).
\end{equation*}
\begin{claim}
For all $j \in \{1, 2, \dots, \npack\}$, $i \in \{1, 2, \dots, n_j\}$ we have
\begin{equation}
\label{clm:packing_intersect}
    \measure{P^j \cap R^j_i} > \frac{\measure{P^j} + \measure{R^j_i} - \beta \e}{2}.
\end{equation}
\end{claim}
\bpf  Fix $j \in \{1, 2, \dots, \npack\}$. For all $i \in \{1, 2, \dots, n_j\}$ we have
\begin{align*}
    \measure{P^j} + \measure{R^j_i} &= \measure{P^j \backslash R^j_i} + \measure{R^j_i \backslash P^j} + 2 \measure{P^j \cap R^j_i} \\
    &= \measure{\sdiff{P^j}{R^j_i}} + 2 \measure{P^j \cap R^j_i} < \beta \e + 2 \measure{P^j \cap R^j_i}.
\end{align*}
The first equality follows from straightforward accounting, and the second from the definition of symmetric difference. The inequality follows from the manner in which set $R^j_i$ is matched to the packing set $P^j$. Finally, a simple rearrangement of terms yields the result. \epf\\
This proves that the intersection of each set $R^j_i$ with its corresponding first-level packing set $P^j$ is heavy. This lets us define a second-level packing using the intersections $R^j_i \cap P^j$.

Rather than directly bounding the the number of processed sets assigned to a  first-level packing set, it is easier to first bound the length of a random subsequence of the assigned sets $\mcS^j$. 
For any $j \in \{1, \dots, \npack\}$, define the subsequence $\mcS'^j$ as the subsequence of processed sets $R$ in $\mcS^j$ whose corresponding samples $H_R$ form $\gamma$-nets for the system $(R, \reduce{\sets}{R})$:
\begin{equation*}
    \mcS'^j = \left(R \in \mcS^j: H_{R} \textrm{ is a } \gamma\textrm{-net for } (R, \reduce{\sets}{R} ) \right).
\end{equation*}
We proceed by bounding the length of the subsequence $\mcS'^j$, and by choosing $\gamma$ so as to make it likely for a set $R$ in $\mcS^j$ to be in $\mcS'^j$, using the $\e$-net \cref{thm:e-net}. We use this to upper bound the expected number of sets in $\mcS^j$. Let $\len{\mcS}$ denote the length of a sequence $\mcS$. 

The following claim bounds the length of the subsequence above.
\begin{claim}
    For any $j \in \{1,2,\dots, \npack\}$:
    \begin{equation}
    \label{clm:subsequnece_length}
        \len{\mcS'^j} \leq
        \begin{cases}
        \frac{24d}{3/2 - \beta - \gamma}\cdot  \cells{\frac{8d}{3/2-\beta-\gamma}, \frac{48d}{3/2-\beta - \gamma}}, & \textrm{ if } \beta + \gamma \geq 1/2; \\
        \bigO{1}, & \textrm{otherwise.}
        \end{cases}
    \end{equation}
\end{claim}
\bpf Let $n'_j = \len{\mcS'^j}$ and relabel the sets so that $\mcS'^j = \left(R^j_1, \dots, R^j_{n'_j}\right)$. Now consider an auxiliary sequence of sets based on intersecting the entries $R^j_i$ in $\mcS'^j$ with $P^j$:
\begin{equation*}
    \mcT'^j = \left(S^j_1, \dots, S^j_{n'_j}\right) \quad \textrm{ with } \quad S^j_i = R^j_{i} \cap P^j \textrm{ for each } i \in \{1, \dots, n'_i\}.
\end{equation*}
This sequence of sets is used to generate a second-level packing. To do this, consider two distinct set-indices $1 \leq k < l \leq n'_j$. The points $H_{R^j_k}$ are added before set $R^j_l$ is considered, so $H_{R^j_k}$ is a $\gamma$-net for $\left(R^j_k, \reduce{\sets}{R^j_{k}}\right)$, whereas the set $R^j_l$ -- because it is subsequently considered by the algorithm  -- is not hit by this net.  This implies that the intersection of $R^j_k$ and $R^j_l$ is of bounded weight, as it would be hit by the $\gamma$-net otherwise:
\begin{equation*}
    \measure{R^j_k \cap R^j_l} < \gamma \cdot \measure{R^j_k}.
\end{equation*}
This implies that the weight of the intersection of $S^j_k$ and $S^j_l$ is bounded above:
\begin{equation}
    \label{eq:bounded_intersect_measure}
    \measure{S^j_k \cap S^j_l} = \measure{R^j_k \cap R^j_l \cap P^j} \leq  \measure{R^j_k \cap R^j_l} < \gamma \cdot \measure{R^j_k}.
\end{equation}

The fact that sets in $\mcT'^j$ have pairwise intersections of small weight implies that their symmetric differences are heavy:  
\begin{align*}
    \measure{\sdiff{S^j_k}{S^j_l}} &= \measure{S^j_k} + \measure{S^j_l} - 2\measure{S^j_k \cap S^j_l} \\
    &= \measure{R^j_k \cap P^j} + \measure{R^j_l \cap P^j} - 2\cdot \measure{S^j_k \cap S^j_l} \\
    &> \frac{\measure{P^j} + \measure{R^j_k} - \beta\e}{2} + \frac{\measure{P^j} + \measure{R^j_l} - \beta\e}{2} - 2\cdot \measure{S^j_k \cap S^j_l} \\
    & > \frac{\measure{P^j} + \measure{R^j_k} - \beta\e}{2} + \frac{\measure{P^j} + \measure{R^j_l} - \beta\e}{2} - 2 \gamma \cdot \measure{R^j_k}\\
    &= \measure{P^j} - \beta \e + \frac{1}{2}\measure{R^j_l} + \left(1/2 - 2 \gamma \right) \measure{R^j_k}\\    
    &\geq \left(3/2 - \beta - \gamma\right) \cdot\measure{P^j},
\end{align*}
where the first inequality uses \cref{clm:packing_intersect}, the second \cref{eq:bounded_intersect_measure}, and the last exploits the fact that sets $R^j_k$, $R^j_l$, and $P^j$ are each of measure at least $\e$ and at most $2\e$, and that $\gamma \leq 1/4$. Thus, depending on the constants, the sequence $\mcT'^j$ may form a weighted packing.

Finally, reviewing two cases for the constants $\beta$ and $\gamma$ makes the above more precise. First, if $\beta + \gamma < 1/2$, the inequality above implies that the symmetric difference of $S^j_k$ and $S^j_l$ is strictly larger than $\measure{P^j}$. This cannot be the case as both sets are subsets of $P^j$. Thus, the only sequence $\mcS'^j$ for which $\beta + \gamma$ can be less than a half is if there are no two unique indices, implying that $\len{\mcS'^j} \leq 1$. Secondly, if $\beta + \gamma \geq 1/2$, the sets in $\mcT'^j$ form a $\left(\measure{P^j}, (3/2 - \beta - \gamma)\measure{P^j}\right)$-packing over $P^j$; all sets have measure at most $\measure{P^j}$, and every symmetric difference is at least $(3/2 - \beta - \gamma)\measure{P^j}$. This is our second-level packing. Now, the Shallow Packing \cref{lem:shallow_packing} implies:
\begin{equation*}
    \len{\mcS'^j} = \len{\mcT'^j} \leq \frac{24d}{3/2 - \beta - \gamma} \cdot \cells{\frac{8d}{3/2 - \beta - \gamma}, \frac{48d}{3/2 - \beta - \gamma}},
\end{equation*}
where we have used the fact that $\cells{\cdot, \cdot}$ is non-decreasing and that $\mu(P^j) \leq 1$. \epf 

We can now bound the length of the full sequence of sets assigned to the packing set $P^j$. Taking expectations sidesteps any dependencies in the sequences. For instance, a set $R$ can only be in $\mcS^j$ if previous samples failed to hit it. However, for each fixed set $R \in \sets$, the probability of the sampled points $H_S$ forming a $\gamma$-net for $(R, \reduce{R}{\sets})$ is independent of previous sampling. Indeed, by \cref{thm:e-net}, the probability that $H_R$ is a $\gamma$-net is at least $1 - \gamma \geq 1/2$.
\begin{lemma}[Mustafa, Lemma 5 \cite{mustafa19}]
\label{lem:expected_additions}
\begin{equation*}
    \Ex\left[\len{\mcS^j} \big\rvert n_j \geq 1 \right] \leq \frac{48}{3/2 - \beta - \gamma} \cdot \cells{\frac{8d}{3/2 - \beta - \gamma}, \frac{48d}{3/2 - \beta - \gamma}} 
\end{equation*}
\end{lemma}
\bpf We use a simple application of linearity of expectation, and \cref{thm:e-net}: 
\begin{equation*}
    \Ex[\len{\mcS'^j} \big\rvert n_j \geq 1] = \sum_{R \in \mcS^j}\Pr[H_R \textrm{ is  a } \gamma\textrm{-net for } (R, \reduce{\sets}{R})] \geq \tfrac{1}{2} \cdot \len{\mcS^j},
\end{equation*}
where we drop the conditioning on $n_j \geq 1$ because the event that a particular sample $H_R$ is a $\gamma$-net is independent of the number of previous samples. On the other hand, \cref{clm:subsequnece_length} upper bounds the size of $\len{\mcS'^j}$. Piecing these together yields the inequality:
\begin{equation*}
    \tfrac{1}{2}\len{\mcS^j} \leq \Ex\left[\len{\mcS'^j} \big\rvert n_j \geq 1\right] \leq \len{S'^j} \leq \frac{24d}{3/2 - \beta - \gamma}\cells{\frac{8d}{3/2-\beta-\gamma}, \frac{48d}{3/2 - \beta - \gamma}}.
\end{equation*}
\epf 

Thus far we have conditioned on a set in the first-level packing being assigned at least one processed set. We now bound the probability of this being the case. Later, this probability is used to compute the expected number of processed sets assigned to a first-level packing set.
\begin{lemma}
\label{lemm:prob_of_process}
Let $H_0$ be the initial sample taken by the algorithm. Then for any $j \in \{1, \dots, \npack\}$:
\begin{equation*}
    \Pr[n_j \geq 1] = \bigO{\frac{1}{d^2 \cells{\tfrac{8d}{\beta \e}, \tfrac{48d}{\beta}}^2}}. 
\end{equation*}
\end{lemma}
\bpf Fix an index $j \in \{1, \dots, \npack\}$. Suppose that $n_j \geq 1$. By \cref{clm:packing_intersect}, for any $i \in \{1, \dots, n_j\}$:
\begin{align*}
    \measure{P^j \cap R^j_i} &> \frac{\measure{P^j} + \measure{R^j_i} - \beta\e}{2} \\
    &\geq \frac{\measure{P^j}+\measure{P^j}/2 - \beta \measure{P^j}}{2} = \left(\frac{3}{4} - \frac{\beta}{2}\right)\cdot \measure{P^j}
\end{align*}
The second inequality above follows from the assumption that all sets have weights within a factor 2 of each other. The above implies that, if $H_0$ is a $\left(\frac{3}{4} - \frac{\beta}{2}\right)$-net for $\left(P^j, \reduce{\sets}{P^j}\right)$, then any $R \in \mcS^j$ would be hit by $H_0$. In other words, $n_j \geq 1$ only if $H_0$ is not a $\left(\frac{3}{4} - \frac{\beta}{2}\right)$-net for $\left(P^j, \reduce{\sets}{P^j}\right)$:
\begin{equation*}
    \Pr[n_j \geq 1] \leq \Pr\left[H_0 \textrm{ is not a } \left(\frac{3}{4} - \frac{\beta}{2}\right)\textrm{-net for 
 } \left(P^j, \reduce{\sets}{P^j}\right)\right].
\end{equation*}
Beacuse $\tfrac{\mu_j}{\e} \geq \tfrac{\mu_j}{\mu(R)}$, the initial sample includes each element with sufficient probability to apply \cref{thm:e-net} to the RHS above, completing the proof.\epf

\subsection{Proof of \cref{thm:algo_with_consts}}
\textit{Proof of \cref{thm:algo_with_consts}}. At this stage, the analysis closely follows Mustafa's \cite{mustafa19}. Clearly, the algorithm proceeds until $H$ is an $\e^*$-net with respect to measure $\mu^*$, i.e., a hitting set. It suffices to bound the expected size of the hitting set $H$, as well as the expected number of oracle calls. These quantities are related, since the number of points added depends on the number of times a set is processed.

First, consider the expected size of the hitting set. There are two contributions to the set: the initial sample $H_0$, and the samples from the processed sets $H_{R_1}, \dots, H_{R_T}$. We bound the expected size of the initial sample first.
\begin{claim}
    
    The expected size of the initial sample, $\Ex[\card{H_0}]$ is bounded by
      \begin{equation}
     \label{clm:n_initial_points}
     \bigO{ \frac{1}{\left(\tfrac{3}{2} - \tfrac{\beta}{2}\right)\e}\max\left\{\log\left(d \cells{\frac{8d}{\beta \e}, \frac{48d}{\beta}}\right), \frac{d}{\left(\tfrac{3}{2} - \tfrac{\beta}{2}\right)}\log\frac{1}{\left(\tfrac{3}{2} - \tfrac{\beta}{2}\right)\e}\right\}}.
 \end{equation}  
\end{claim}
This follows by summing the probability of sampling $x$ for each $x \in X$. An analogous result is used for the number of points added during the processing of a set $R \in \sets$, provided it is processed:
\begin{claim}
    For any fixed set $R \in \sets$, conditional on being processed, the expected number of points added each time it is processed is
    \begin{equation}
    \label{clm:n_added_points}
      \Ex\left[\card{H_R} \right] \leq 2 \left(\frac{\log 2}{\gamma} + \frac{d}{\gamma}\log\frac{1}{\gamma}\right) = \bigO{1}.
    \end{equation}
\end{claim}
This bound applies irrespective of whether or not a set was processed previously. 

The number of points added during processing, and the number of oracle calls, can be bounded together. Recalling that $R_1, \dots, R_T$ are the processed sets, and using the claim above, the number of added elements is at most
\begin{equation}
    \label{eqn:expected_added_points}
    \Ex\left[\sum^T_{i=1}\card{H_{R_i}}\right] \leq \Ex\left[\sum^T_{i=1}2\left(\frac{\log 2}{\gamma} + \frac{d}{\gamma}\log\frac{1}{\gamma}\right)\right]  =  \Ex[T] \cdot 2\left(\frac{\log 2}{\gamma} + \frac{d}{\gamma}\log \frac{1}{\gamma}\right).
\end{equation}
Thus, it suffices to bound the expected number of oracle calls $\Ex[T]$. This is where we employ both the first-, and second-level packings. In particular
\begin{equation}
\label{eq:runtime_breakdown}
    \Ex[T] = \Ex\left[\sum^{\npack}_{j = 1}\len{\mcS^j}\right] = \underbrace{\sum^p_{j=1}}_{\textrm{(i)}} \cdot \ \underbrace{\Ex[\len{\mcS^j} \big\rvert n_j \geq 1 ]}_{\textrm{(ii)}} \ \cdot\  \underbrace{\Pr[n_j \geq 1]}_{\textrm{(iii)}} .
\end{equation}
The terms (i), (ii) and (iii) are bounded using \cref{eq:size_pack}, \cref{lem:expected_additions}, and \cref{lemm:prob_of_process}, respectively. In addition, using $\tfrac{3}{2} - \beta - \gamma \geq \tfrac{1}{2} \geq \max\{\beta\e, \beta/2\}$:
\begin{enumerate}[(i)]
    \item $\npack \leq \frac{24d}{\beta \e} \cells{\frac{8d}{\beta\e}, \frac{24d}{\beta}}$ 
    \item $\Ex\left[\len{\mcS^j} \big\rvert n_j \geq 1 \right] \leq \frac{48d}{3/2 - \beta - \gamma} \cells{\frac{8d}{3/2 - \beta - \gamma}, \frac{24d}{3/2 - \beta - \gamma}} \leq \frac{48d}{3/2 - \beta - \gamma} \cells{\frac{8d}{\beta\e}, \frac{24d}{\beta}}$;
    \item $\Pr[n_j \geq 1] \leq \left(d^2 \cells{\frac{8d}{\beta \e}, \frac{24d}{\beta}}^2 \right)^{-1}$.
\end{enumerate}
Combining the right-hand-side terms, we obtain the bound
\begin{equation}
    \label{eq:bound_steps}
    \Ex[T] \leq \frac{24 \cdot 48}{\beta \cdot (3/2 - \beta - \gamma)}\frac{1}{\e} = \bigO{\frac{1}{\e}}.
\end{equation}
This is minimized by choosing a small $\gamma$, e.g. $\gamma = 1/ 100$, and setting $\beta = 3/4$. 

Finally, summing over the $\ell$ groups of sets, and adding the expected number of initial samples to the expected  number added points completes the proof. Note that \cref{eq:bound_steps} also bounds the expected number of oracle calls made during the run of the algorithm.  \epf

%
%
%
\bibliographystyle{splncs04}
\bibliography{references.bib}
%

\input{appendix}

\end{document}

%% file: appendix.tex
\begin{appendix}
\newpage
\section{Proof of the Weighted Packing Lemma}
\label{app:WeightedPackingLemma}

This section proves the Weighted Packing Lemma \ref{lem:packing}. Our proof closely follows the original in Haussler \cite{Haussler1995ShperePacking}. The reader is referred to Matou\v{s}ek \cite{Matousek2009GeometricDA} (Sec 5.3) for an excellent treatment of the unweighted proof. We begin by restating the weighted lemma.

\begin{lemma}[Weighted Packing Lemma]
Let $(X, \pack)$ be a  set system with $\neles$ elements and $\nsets$ sets $\pack = \{S_1, \dots, S_n\}$, equipped with weights $\mu:X \rightarrow \mathbb{R}_{\geq 0}$ with $\mu(X) > 0$. Let $\vcdim(\pack) \leq d$ for some integer $d \geq 1$, and let $\delta > 0$ be a constant such that $\mu(\Delta(S_i, S_j)) \geq \delta $ for all $1 \leq i < j \leq \nsets$. Then
\begin{equation*}
    \card{\pack } \leq 2 \Ex \left[\card{\reduce{\pack}{Y} }\right]
\end{equation*}
where $Y$ is the set of unique elements in an random sample $U = (u_1, u_2, \dots, u_s)$ of size $s = \lceil \frac{8d}{\delta}\rceil -1$, in which each element $u_k$  is sampled iid $u_k \sim \mu(\cdot)$ with replacement. 
\end{lemma}

The proof strategy is the following. First, we consider a random sample $U = (u_1, \dots, u_s)$ sampled from $\mu(\cdot)$ with replacement. The sample $U$ may contain repeated elements; let $Y \subseteq X$ denote the set of unique elements in $U$. Next, we generate a unit-distance graph; a weighted graph that depends on the random set $Y$, and derive three claims about the total weight of its edges: (i) an upper bound on the total weight, (ii) a partial lower bound, and (iii) a complete lower bound. Combining the bounds completes the proof. The main differences between our approach and that of Chazelle, Haussler, and Mustafa are twofold \cite{Chazelle92,Haussler1995ShperePacking,Mustafa16}. Firstly, we permit non-uniform weights $\mu$ as opposed to weighing each set by its cardinality, \textit{and} for we sample from a probability distribution proportional to $\mu$. Secondly we use sampling \textit{with replacement} as opposed to \textit{without replacement}. This makes the analysis more straightforward under non-uniform sampling using $\mu$. 

A weighted unit-distance graph over the sampled set-system takes a central stage in the proof. In this graph, sets are viewed as vertices, and edges are drawn between any two sets at unit-distance of each other.
\begin{definition}[Unit distance graph]
Let $(X, \pack)$ be a set system. The unit distance graph of $(X, \pack$) is a graph $G = (\pack, E_\pack)$ with vertex set $\pack$ and edges
\begin{equation*}
    E_\pack = \{ \{S_i, S_j\} \in \pack \times \pack : \card{\sdiff{S_i}{S_j}} = 1 \text{ and } i \neq j \}
\end{equation*}
\end{definition}
In other words, edges represents pairs of sets that differ on exactly one element. The following lemma connects the number of edges with the VC-dimension.
\begin{lemma}[Haussler \cite{Haussler1995ShperePacking}]
\label{lemma:BoundedEdgeNr}
Fix a set system $(X, \pack)$ with $\vcdim(\pack) = d$. Let $G = (\pack, E_\pack)$ be its unit distance graph. Then $\card{E_\pack } \leq d \card{\pack}$.
\end{lemma}
We use this lemma to bound the total edge weight in our unit-distance graph. But first, we define our particular unit-distance graph and its edge weights.

We construct a graph that depends on the random set $Y \subseteq X$. Consider the projection of $\pack$ to $Y$, denoted $\reduce{\pack}{Y}$. Let $G_Y = (\reduce{\pack}{Y}, E_{\reduce{\pack}{Y}})$ be a unit distance graph over the projected system. For each vertex $S' \in \reduce{\pack}{Y}$,  define the \textit{vertex weight} as the number of sets $S \in \pack$ that are projected to $S'$ in $\reduce{\pack}{Y}$, that is
\begin{equation*}
    w(S') = \card{ \{S \in \pack : S \cap Y  = S'\}}
\end{equation*}
Moreover, define the \textit{weight of an edge} $\{S'_i, S'_j\} \in E_{\reduce{\pack}{Y}}$ as the minimum over the weights of its two vertices, $w\left(\{S'_i, S'_j\}\right) = \min\{w(S'_i), w(S'_j)\}$. Finally, let the \textit{total edge weight} be $W = \sum_{e \in E_{\reduce{\pack}{Y}}}w(e)$. Note that the weights are random variables because they depend on the random selection $Y$. We proceed by bounding the total edge weight. This will allow us to bound the size of the packing in a way that ``looks like a magician's trick'' \cite{Matousek2009GeometricDA}. 

First, we find an upper bound for the total edge weight. This is naturally also an upper bound on the expected total edge weight.
\begin{claim}
\label{claim:final_ub}
The total edge weight is upper-bounded by $W \leq 2d\card{\pack}$.
\end{claim}
\textit{Proof.} This is the proof of Haussler and Chazelle \cite{Haussler1995ShperePacking,Chazelle92}. \cref{lemma:BoundedEdgeNr} implies that:
\begin{equation*}
    \sum_{S' \in \reduce{\pack}{Y}}\deg(S') = 2\card{E_{\reduce{\pack}{Y}}} \leq 2d \card{\reduce{\pack}{Y}}.
\end{equation*}
Hence there exists a vertex $S'$ in $\reduce{\pack}{Y}$ with degree at most $2d$. Each edge incident to $S'$ has weight at most $w(S')$, so the vertex $S'$ is responsible for edges of total weight at most $2dw(S')$. Applying this inductively proves the bound on the total weight.
\begin{equation*}
    W \leq \sum_{S' \in \reduce{\pack}{Y}}(\textrm{Weight due to } S') \leq 2d \sum_{S' \in \reduce{\pack}{Y}}w(S') = 2d\card{\pack} = 2d\nsets
\end{equation*}
\epf 

Next we derive a lower bound. It suffices to derive the bound for a reduced problem. Let $U'$ be the subsequence of sample $U = (u_1, \dots, u_s)$ containing all but the last element $u_s$. Similarly, let $Y'$ be the set of unique elements in the subsample $U'$, just as $Y$ denotes the unique elements in the full sample $U$. Now, when the final element $u_s$ is added to $U'$, some vertices in $\reduce{\pack}{Y}$ may form an edge \textit{due to} $u_s$. This occurs exactly when the last sampled element $u_s$ falls in the symmetric difference of two sets that were previously equal in the projection $\reduce{\pack}{Y'}$. Let $W_s$ be the sum of the weights of the edges due to element $u_s$. In other words, the weight $W_s$ is the weight of the edges generated by adding a random element $u_s \sim \mu(\cdot)$ to the random sequence $U' = (u_1, \dots, u_{s-1})$. Because the samples are iid, given a squence, the \textit{ordering} of the random elements in the sequence is uniform over all permutations, so we can apply symmetry of expectation:
\begin{equation*}
    \Ex[W] = s\Ex[W_s].
\end{equation*}
Hence, it suffices to derive an upper bound on $\Ex[W_s]$.

Following Haussler \cite{Matousek2009GeometricDA}, an intermediate step is to lower bound the expectation of the weight $W_s$ due to the $s$th element $u_s$ conditional on the previous $s-1$ elements $U'$. The lower bound follows from considering pairs of vertices that lack edges in $E_{\reduce{\pack}{Y'}}$ but that may share an edge after adding $u_s$. This happens exactly when the new element $u_s$ falls in their symmetric difference. This event occurs with probability at least $\delta$, because we assume $\mu(\sdiff{S}{}) \geq \delta$ for all pairs $S \neq R$ in our packing $\pack$. This argument uses the fact that we sample elements proportional to the weight $\mu$.
\begin{claim}
    \label{claim:partial_lb}
    For any set $Y' \subseteq X$ generated by a fixed sequence $u' = (u_1, \dots, u_{s-1})$, with $u_k \in X$ for $k = 1, \dots, s- 1$, it holds that
    \begin{equation*}
        \Ex_{\mu}[W_s \mid U' = u'] \geq \frac{\delta}{2}\left(\nsets - \card{\reduce{\pack}{Y'}}\right)
    \end{equation*}
where $\Ex_\mu$ denotes expectation over is over a single random element $u_s \sim \mu(\cdot)$.
\end{claim}
\textit{Proof}. Fix a set $Y' \subseteq X$ corresponding to the unique elements in the partial selection $u' = (u_1, \dots, u_{s-1})$. Consider an arbitrary set $Q $ in the projection $\reduce{\pack}{Y'}$. There may be many sets in $\pack$ that map to $Q$ in $\reduce{\pack}{Y'}$. Let $\pack_Q$ be the collection of these sets, and let $b$ denote the number of such sets. Note that for any pair of sets $S_i, S_j \in \pack_Q$, $Y'$ cannot contain any element in the symmetric difference $\Delta(S_i, S_j)$, or else the two sets would not map to the same $Q$. However, when an additional element $u_s$ is sampled and added to $Y'$ (with the possibility that $u_s$ is already in $Y'$) the collection $\pack_Q$ is partitioned into two groups: (i) sets that contain $u_s$, and (ii) sets that do not contain $u_s$. Let $b_1$ and $b_2$ denote the number of sets in these groups, respectively, with $b = b_1 + b_2$. These two groups are at a unit-distance in $\reduce{\pack}{Y}$. The weight of the resulting edges in the unit-distance graph is exactly $\min\{b_1, b_2\}$.

By adding up the expected weights due to each pair we get $\Ex[W_s]$. For every pair $S'_1, S'_2 \in \pack_Q$, the probability that $u_s$ hits $\Delta(S'_1, S'_2)$ is $\mu(\sdiff{S'_1}{S'_2}) \geq \delta$. Thus, the expected contribution of each pair to the product $b_1 b_2$ is at least $\delta$. Note that the sum $b_1 + b_2 = b$ depends on $Y'$, however is independent of $u_s$. Hence
\begin{align*}
    \Ex[\min\{b_1, b_2\}] &\geq \Ex\left[\frac{b_1 b_2}{b}\right] \\
&= \sum_{S'_1, S'_2 \in \pack_Q}\frac{\Pr\left[U_s \in \sdiff{S'_1, S'_2}\right]}{b} \\
&= \frac{\card{\pack_Q}\left(\card{\pack_Q} - 1\right)\delta}{2\card{\pack_Q}} = \frac{\delta}{2}\left(\card{\pack_Q} - 1\right).
\end{align*}
The first inequality follows from the fact that $\min\{b_1, b_2\} \geq b_1b_2 / b$. The first equality uses the fact that there $\sets_Q$ is partitioned into two groups, each at unit distance, so $b_1 b_2$ is the total number of edges between the two groups.
Taking the above inequality and summing over all vertices $Q \in \reduce{\pack}{Y'}$ gives the result
\begin{align*}
    \Ex[W_s \mid U' = u'] &\geq \sum_{Q \in \reduce{\pack}{Y'}}\frac{\delta}{2}\left(\card{\pack_Q} - 1\right)\\
    &= \frac{\delta}{2}\left(\sum_{Q \in \reduce{\pack}{Y'}}\card{\pack_Q} - \sum_{Q \in \reduce{\pack}{Y'}}1\right) = \frac{\delta}{2}\left(\nsets - \card{\reduce{\pack}{Y'}}\right).
\end{align*}
\epf 

By using the claim above it is now straightforward to produce a lower bound on the expected total edge weight.
\begin{claim}
    \label{claim:final_lb}
    $\Ex[W] \geq 4d\nsets - 4d\Ex[\card{\reduce{\pack}{Y'}}]$.
\end{claim}
\textit{Proof.} We employ the reduction from above as well as the partial lower bound. Let $X^k$ denote the Cartesian product of element set $X$. For any length sequence $u' = (u_1, \dots, u_{s-1})$ of elements in $X^{s-1}$, let $Y'(u') \subseteq X$ be the set of unique elements in $u'$. Analogously let let $Y'(U')$ denote the random set of unique elements in a random sequence $U'$ in $X^{s-1}$. It then follows that:
\begin{align*}
    \Ex[W] &= s \Ex[W_s] \\
    &= s \sum_{u' \in X^{s-1}}\Ex[W_s \mid Y'(U') = Y(u')] \cdot \Pr[U' = u']\\
    &\geq s \sum_{u' \in X^{s-1}}\frac{\delta}{2}\left(\nsets  - \card{\reduce{\pack}{Y(u')}}\right)\cdot \Pr[U' = u']\\
    &= \frac{s \delta}{2}\left(\nsets \sum_{u' \in X^{s-1}}\Pr[U' = u'] \ - \sum_{u' \in X^{s-1}}\card{\reduce{\pack}{Y(u')}}\cdot \Pr[U' = u']\right) \\
    &= \frac{s \delta}{2} \left(\nsets - \Ex\left[\card{\reduce{\pack}{Y'(U')}}\right]\right),
\end{align*}
where the first inequality uses the partial lower bound, and the last equality follows from the definition the random set $Y(U')$. \epf

Finally, all the pieces are in place to prove the Packing Lemma. Using the lower and upper bounds it follows that cardinality of $\pack$ is bounded above by twice the expected size of $\reduce{\pack}{Y'}$:
\begin{align*}
    4d\nsets - 4d\Ex[\card{\reduce{\pack}{Y'}}] \leq \Ex[W] \leq 2d\nsets.
\end{align*}
This yields the statement of the weighted packing lemma, completing the proof.
\epf

\end{appendix}